\documentclass[aps,showpacs,twocolumn,prl]{revtex4}
\usepackage{amssymb}
\usepackage{natbib}
\usepackage{amsmath}
\usepackage{amsfonts}
\usepackage{graphicx}
\usepackage{mathrsfs}
\usepackage{dcolumn}
\usepackage{bm}
\usepackage{color}
\definecolor{rot}{rgb}{0.75,0.05,0.25}
\definecolor{hellgrau}{gray}{0.5}
\definecolor{blau}{rgb}{0,0,0.7}

\definecolor{rot}{rgb}{0.75,0.05,0.25}
\definecolor{hellgrau}{gray}{0.5}
\definecolor{blau}{rgb}{0,0,0.7}

\begin{document}

\title{Comment on ``Tsallis power laws and finite baths with negative heat capacity"}
\author{Michele Campisi}\email{Michele.Campisi@physik.uni-augsburg.de}
\address{Institute of Physics, University of Augsburg, Universit\"atsstrasse 1, D-86153 Augsburg, Germany}
\date{\today }
\begin{abstract}
In [Phys. Rev. E 88, 042126 (2013)] it is stated that Tsallis distributions do not emerge from thermalization with a ``bath''
of finite, energy-independent, heat capacity. We report evidence for the contrary.
\end{abstract}
\maketitle

In the abstract of Ref. \cite{Bagci13PRE88} the authors state that
``Tsallis distributions with fat tails are possible only for finite baths with constant negative heat capacity, while constant positive heat capacity finite baths yield decays with sharp cutoff with no fat tails''. In the conclusions, instead, they appear to argue for the contrary: ``The most important question is finally to decide whether the Tsallis distributions are indeed to emerge from the coupling of the physical system with a finite bath. The answer to this is that they do not since the emergence of the Tsallis distributions in the finite bath scenario, be it ordinary or escort, requires the constant heat capacity of the finite heat bath to be $q$ dependent."

This confusing state of affairs calls for a due clarification.

If a classical system (S) stays in weak contact with a ``finite bath'' (FB), i.e., a system having \emph{constant} (i.e., energy-independent) \emph{finite heat capacity} $C$, then, provided the dynamics of the S+FB compound is ergodic, the marginal distribution of S is given by
the following two expressions \cite{Campisi09PRE80,Campisi12EPL99}:
\begin{eqnarray}
 p_>(\mathbf{x},\mathbf{p})=\frac{[E_{\text{tot}}-H_S(\mathbf{x},\mathbf{p})]^{C-1}}
 {\int \mathrm{d}\mathbf{x}\mathrm{d}\mathbf{p}[E_{\text{tot}}-H_S(\mathbf{x},\mathbf{p})]^{C-1}},\quad C>0 \,
\label{eq:power-law2}\\
 p_<(\mathbf{x},\mathbf{p})=\frac{[H_S(\mathbf{x},\mathbf{p})-E_{\text{tot}}]^{C-1}}
 {\int \mathrm{d}\mathbf{x}\mathrm{d}\mathbf{p} [H_S(\mathbf{x},\mathbf{p})-E_{\text{tot}}]^{C-1}}, \quad C<0 \, ,
\label{eq:power-law}
\end{eqnarray}
depending on whether the heat capacity of the FB is positive or negative. Here $E_\text{tot}$ is the energy of the compound $S+FB$,
$H_S(\mathbf{x},\mathbf{p})$ is the S Hamiltonian, and for simplicity we have set $k_B=1$.

Expressions (\ref{eq:power-law2},\ref{eq:power-law}) follow by integrating the FB degrees of freedom out of the microcanonical distribution of energy $E_\text{tot}$ of the S+FB compound \cite{Campisi09PRE80,Campisi12EPL99}, according to the basic rules
of probability theory.
An example of a system with a constant positive heat capacity is given by a set of $N$ hard spheres in a $d$-dimensional box.
In this case the heat capacity is $C=dN/2$. An example of a system with constant negative heat capacity is given by 
a single particle in the $1/r$ potential in 3-dimensional space. In this case the heat capacity is $C=-2/3$.
The predictions of Eqs. (\ref{eq:power-law2},\ref{eq:power-law}) have been excellently corroborated by numerical simulations with a FB  made of 1,2,3 and 4 hard disks in 2D  \cite{Campisi09PRE80}, and made of a single particle in the $1/r$ potential \cite{Campisi12EPL99}.

Following Ref. \cite{Campisi07PLA366}  Eqs.  (\ref{eq:power-law2},\ref{eq:power-law}) can be equivalently compactly re-written as:
\begin{eqnarray}
p(\mathbf{x},\mathbf{p})  = \left[ 1 - \beta(1-q)(H_S(\textbf{x},\textbf{p})-U)\right]^{\frac{q}{1-q}}/N 
\label{eq:power-lawb}
\end{eqnarray}
Where $C=(1-q)^{-1}$, $U$ is the average energy of S, $N$ is the normalizations, and $\beta^{-1}= (E_{\text{tot}}-U)/C$ is the bath's temperature (namely twice its average kinetic energy per degree of freedom). The cases $C > 0, C<0$, correspond to $q<1, q>1$, respectively.

The distributions in (\ref{eq:power-lawb}) are the Tsallis escort distributions, with sharp cut-off ($q<1$)
and long tail ($q>1$).

A point that the authors of Ref. \cite{Bagci13PRE88} seem not to appreciate (indeed it seems to be more a source of concern for them)
is that there is a  simple relation between $C$ and $q$. This offers an interpretation of the physical meaning of the otherwise elusive $q$, in the particular scenario of a system in weak contact with a finite bath: $q$ contains information about a central thermodynamic measurable property of the FB, namely its heat capacity $C$. That is, here $q$ is not a mere fitting parameter as customary in the non-extensive thermodynamic literature, but is instead a quantity that can be independently measured and predicted from knowing the Hamiltonian of the FB. 

As reported in Ref. \cite{Campisi07PLA362} by one of the authors of the here commented paper \cite{Bagci13PRE88} and myself, the following equipartition theorem holds for the distributions in (\ref{eq:power-lawb}):
\begin{eqnarray}
\left\langle p_i\frac{\partial H}{\partial p_i}\right\rangle = \frac{1}{\beta}
\label{eq:equipart}
\end{eqnarray}
where repeated indexes are not summed, and $\langle \cdot \rangle$ denotes averaging over $p(\mathbf{x},\mathbf{p})$.
This mathematical theorem contrasts starkly with the statement appearing in the abstract of Ref. \cite{Bagci13PRE88}: ``the correspondence between Tsallis distributions and finite baths holds at the expense of violating the equipartition theorem for finite classical systems at equilibrium'' \footnote{A simple way to prove Eq. (\ref{eq:equipart}) is as follows. Consider the total compound $H_\text{tot}=H_S+H_{FB}+h$ in the microcanonical ensemble of energy $E_\text{tot}$. $H_{FB}$ and $h$ are the FB Hamiltonian and the small interaction Hamiltonian, respectively. According to the classical microcanonical equipartition 
theorem it is $\langle z_i \partial H_\text{tot} /\partial z_i\rangle_{E_\text{tot}} = T(E_\text{tot})$, where $z_i$ is any of the canonical variables of the S+FB compound, $T$ is temperature and $\langle \cdot \rangle_{E_\text{tot}}$ denotes microcanonical average. Now call $p_i$ any momentum of S, and $P_i$ any momentum of FB. Neglecting the small term $h$, we have 
\[
\left \langle p_i \frac{\partial H_S}{\partial p_i}\right\rangle_{E_\text{tot}} = \left\langle P_i \frac{\partial H_{FB}}{\partial P_i}\right\rangle_{E_\text{tot}}
\]
 The r.h.s of this equation is what we have called $1/\beta$ in the main text. By performing the integration over the FB variables in the l.h.s, and remembering that $p(\mathbf{x},\mathbf{p})$ in Eq. (\ref{eq:power-lawb}) is the marginal distribution obtained by integrating  the FB degrees of freedom out of the total microcanonical ensembles \cite{Campisi09PRE80,Campisi12EPL99}, the l.h.s can be equivalently written as the average of $p_i \partial H_S /\partial p_i$ over the Tsallis distribution in Eq. (\ref{eq:power-lawb}). Hence Eq. (\ref{eq:equipart}).}.

In sum, it is an incontrovertible fact that the Tsallis escort distributions emerge as the equilibrium distribution of systems in contact with finite baths having (either  positive or  negative) energy-independent heat capacities, and that the equipartition theorem holds in these cases.

\end{document}